\documentclass[conference]{IEEEtran}
\IEEEoverridecommandlockouts
\usepackage{cite}
\usepackage{amsmath,amssymb,amsfonts}
\usepackage{algorithmic}
\usepackage{graphicx}
\usepackage{textcomp}
\usepackage{xcolor}
\def\BibTeX{{\rm B\kern-.05em{\sc i\kern-.025em b}\kern-.08em
    T\kern-.1667em\lower.7ex\hbox{E}\kern-.125emX}}
\begin{document}

\makeatletter
\newcommand{\linebreakand}{%
  \end{@IEEEauthorhalign}
  \hfill\mbox{}\par
  \mbox{}\hfill\begin{@IEEEauthorhalign}
}
\makeatother

\title{Why Teach Quantum?: Elementary Teachers Initial Beliefs about Quantum \\
\thanks{This work was funded through the NSF EAGER: Quantum Teaching and Learning in Elementary Classrooms, award 2329874.}
}

\author{\IEEEauthorblockN{Xiaolu Zhang}
\IEEEauthorblockA{\textit{School of Education} \\
\textit{George Mason University}\\
Fairfax, Virginia, U.S. \\
xzhang22@gmu.edu}
\and
\IEEEauthorblockN{Nancy Holincheck}
\IEEEauthorblockA{\textit{School of Education} \\
\textit{  George Mason University  }\\
Fairfax, Virginia, U.S.\\
nholinch@gmu.edu}
\and
\IEEEauthorblockN{Jessica L. Rosenberg}
\IEEEauthorblockA{\textit{Dept. of Physics and Astronomy} \\
\textit{George Mason University}\\
Fairfax, Virginia, U.S.\\
jrosenb4@gmu.edu}
\linebreakand 

\IEEEauthorblockN{Stephanie Dodman}
\IEEEauthorblockA{\textit{School of Education} \\
\textit{George Mason University)}\\
Fairfax, Virginia, U.S. \\
sdodman@gmu.edu}
\and
\IEEEauthorblockN{Ben Dreyfus}
\IEEEauthorblockA{\textit{Dept. of Physics and Astronomy} \\
\textit{George Mason University}\\
Fairfax, Virginia, U.S. \\
bdreyfu2@gmu.edu}
\and
\IEEEauthorblockN{Jennifer Simons}
\IEEEauthorblockA{\textit{College of Education} \\
\textit{George Mason University}\\
Fairfax, Virginia, U.S. \\
jsimons2@gmu.edu}
}

\maketitle

\begin{abstract}
This paper examines the initial beliefs held by elementary educators (n=11) and their students about teaching and learning quantum concepts at the elementary level. All teachers were participants in a grant-funded project focused on developing teachers’ quantum content knowledge and creating curricular resources to use in elementary classrooms.  Although elementary teachers had limited knowledge of quantum at the beginning of the project, they expressed excitement and a belief that learning quantum would create future possibilities for their students. 
\end{abstract}


\section{Introduction}
Quantum education research has traditionally focused on undergraduate student learning in physics courses \cite{b1, b2, b3}. At the K-12 level, research has predominantly focused on high school physics, with some additional research on computer science applications of quantum \cite{b4,b5}. In recent years, efforts have been made to connect quantum concepts to a variety of high school disciplines and in middle school \cite{b6}.  Our research team is working to introduce quantum to students at an even earlier age, beginning in elementary school. Bringing quantum science to students in the early grades is critical, as children begin to develop their interests and career aspirations during their elementary years \cite{b7}, and they are best positioned to develop positive attitudes towards STEM disciplines before they enter middle school \cite{b8}. Our two-year project examines elementary teacher learning and curriculum development through an extended professional learning program during year 1, as well as classroom implementation and elementary student learning in year 2. This paper focuses on year 1 data and examines the perceptions of 11 elementary educators about quantum teaching and learning. 

\section{Methods}

Design-based Implementation Research (DBIR)\cite{b9} was used as an overarching framework for program design and development. DBIR is particularly appropriate because it focuses on problems of practice from multiple perspectives, is expected to evolve iteratively over the course of the grant, and focuses on developing the capacity for sustaining change in teaching and teacher education \cite{b10}. As a model that emphasizes iterative cycles of design and testing and simultaneous attention to multiple levels of systems in design, DBIR provided a framework for our team to inquire about how to best educate elementary teachers about quantum so that they can learn to integrate quantum in their teaching and develop quantum curricular resources that are relevant to their elementary classrooms. The following research question directed our inquiry: What are teachers’ perceptions about introducing quantum concepts in elementary classrooms?  

\subsection{Context of the Study}

In year 1 of the project (2023-24), we led teachers in twelve sessions of professional learning and curricular resource development. Teachers participated in three in-person sessions on campus with the research team and nine online synchronous sessions using Zoom. Early sessions focused on discussing quantum concepts, including atomic scale and structure, properties of electromagnetic waves, wave-particle duality, quantum superposition, quantum entanglement, applications of quantum, and potential future careers in quantum. 

Recommendations from prior research for integrating quantum into K-12 education \cite{b11, b12} have focused on how to help learners make sense of this challenging and often abstract topic.  Strategies to do this include using visuals and hands-on approaches in teaching quantum, using analogies to make abstract concepts more accessible, focusing more on concepts than mathematics, and aligning quantum with existing educational standards.  We designed our teachers' professional learning experiences around these strategies, using a spiraling method to return to concepts over subsequent professional development sessions. In our PD we used interactive Google Slides and small group discussions around the concepts to support and assess teacher learning.

Elementary teachers in the United States are subject-matter generalists who may have only a cursory understanding of some STEM concepts and often lack confidence in their ability to teach STEM \cite{b13}. To address this, we engaged teachers in different ways to help them learn about quantum and to reflect on their learning.  We provided books related to quantum to our teachers and compiled online resources for them to refer to at key points during professional learning. We were purposeful in surfacing misconceptions about atomic structures and quantum concepts and engaging teachers in discussions and activities to address these misconceptions. 

Beginning in the fifth session of professional development, our teachers began working in small groups to identify small classroom-based inquiries they could engage in to explore the questions they had about their students' learning of quantum and to design and develop new quantum curricular materials. Teachers were encouraged to design small activities that fit within their curriculum and to make discrete connections to quantum concepts.

\subsection{Participants}
During the first year, the research team recruited elementary teachers from schools and districts in Northern Virginia, Washington, D.C., and Maryland. A total of 40 teachers applied to participate, only 31 of whom were elementary teachers. The research team reviewed the applications and selected 11 teachers to participate in the project.  In our selection, we prioritized teacher teams from the same school and aimed to include teachers from multiple school districts.  When choosing between teachers from the same school district, we prioritized teachers from Title I and majority-minority schools.  Our 11 participants included elementary school teachers and STEM or STEAM specialist teachers. These participants were from eight different schools spanning five counties. Five of these schools are Title I schools. The researchers facilitated bi-weekly meetings, amounting to a total of 12 sessions over a six-month period. The structure of these sessions was diverse, blending in-person, virtual meetings via Zoom, and one-on-one check-ins. The in-person meetings were designed to allow participants to engage in hands-on activities that illustrated quantum principles. Meanwhile, the Zoom meetings aimed to clarify quantum concepts and explore how these ideas align with existing curriculum standards. The individual check-ins served as a more private forum, offering participants a chance to voice questions or concerns they might have been hesitant to share in the larger group settings.

\subsection{Data Sources}
Before the initial session, the research team conducted virtual interviews with each participant. These interviews were intended to gather baseline data on the participants’ understanding of quantum science, any connections they perceived with current curriculum standards, their views on STEM education, and their self-perception as STEM educators. The interviews were all recorded on video for thorough analysis. Alongside the interviews, the participants were asked to fill out a survey. This survey aimed to draw out the participants’ questions about quantum concepts and the nuances of integrating these ideas into elementary-level teaching. Additionally, each participant was asked to maintain a journal throughout the study. The journal served as a personal log for them to document new insights, 'Aha' moments, misconceptions regarding quantum, and their emotional journey during the research. The researchers also encouraged participants to prompt questions about quantum with their students, such as ``what is quantum?"  This paper includes data derived from the interviews, surveys, and journal entries of the participants. Institutional Review Board approval was obtained before the study began, and all participants consented to the use of their data in this project. 
     
\subsection{Data Analysis}
Data was qualitatively analyzed to inform real-time changes in the structure of the quantum professional development and to support the ongoing development of curricular materials. The researchers examined participant data throughout the project and engaged in iterative data analysis. As research questions evolved, members of the research team coded data thematically. 

\section{Findings}
The preliminary findings discussed in this section highlight the initial understanding of quantum concepts among our elementary school educators and the baseline knowledge their students possess about quantum science before participating in any structured in-school teaching. We delve into the educators' early stages of familiarity with quantum concepts, exploring how they perceive these ideas both in terms of their own professional learning and their potential implementation in the classroom. This examination includes an assessment of the resources and supports educators believe they would need to effectively teach these advanced scientific topics to young learners. Additionally, the findings regarding students' knowledge of quantum science before any instruction on the topic was provided enable us to pinpoint the conceptual gaps and initial levels of curiosity among the students. Through these findings and the future work of this project, we aim to map out the landscape of quantum education at the elementary level, setting the stage for targeted educational strategies and curricular development that could significantly enhance both teaching and learning outcomes in this cutting-edge area of science.

\subsection{Participants’ Perspectives on Introducing Quantum Concepts to Elementary Students}

A recurring theme among all participants was the belief that quantum science represents the future and that introducing quantum concepts at a young age can create more possibilities for students. Many educators believe that early engagement is particularly vital for students who are at risk of drifting away from the STEM pipeline, often due to limited access to scientific learning opportunities. This point underscores the potential of quantum education to democratize future scientific professions and inspire a broader demographic of students. A participant mentioned, ``I want them to see, like, you choose to [work in a restaurant] because it makes you happy, not because that's the only thing you think is available to you.” She went on to explain that career education is important in her classroom, particularly for fields like quantum, ``I think being able to introduce [students] to people who actually do these jobs and see what the jobs are...to know that the sky is the limit. Talking to them about `you are going to get to dream up jobs that we don't even have in existence yet.'” 
          
Although all participants were enthusiastic about introducing quantum concepts in elementary schools, they also expressed concerns about how to teach them effectively. A survey aimed at uncovering their thoughts about teaching quantum science revealed two main areas of concern: (1) understanding the fundamentals of quantum physics and (2) devising effective methods to teach these concepts to young learners.

Participants posed questions about the fundamentals of quantum physics and its applications. These included: ``Does 'quantum' basically mean very small atomic particles?" ``How can we use quantum science to understand our natural world?" ``How does quantum technology map a magnetic field?"  ``What are some of the major principles of quantum mechanics/theory?" and ``How are quantum computers different from 'normal' computers?" These questions reflect a keen interest in understanding not just the theoretical underpinnings of quantum physics but also its practical implications in technology and everyday life.

Teachers expressed a need for strategies to simplify complex quantum theories for young learners. For instance, one teacher referenced the 2023 National Quantum Day video with LeVar Burton explaining quantum science as the study of very small systems, yet sought further clarification: ``That makes a lot of sense to me, but I am still curious about additional ways to explain it in very simplistic terms. What are some examples of these 'small systems' that might resonate with elementary students?" As another teacher tried to conceptualize quantum, they asked: ``Really small and really cold, are there synonyms we can put in place of quantum to help us define it for ourselves and our students? For example, in describing quantum sensors, is there a better way to say it than really cold, really small sensors? Is it fair to describe it as sensors that can detect information to extremely small sizes and extremely cold temperatures?" Furthermore, participants questioned how to integrate quantum concepts with existing curricula: ``How can we as elementary instructional leaders teach students about quantum and its practical applications?" ``What are some resources I might explore to start to deepen my own knowledge and understanding to be able to teach students?" ``how can I get just a regular old classroom teacher to be excited about quantum and teaching it.”
            
These inquiries indicate that participants are eager not only to acquire a deeper understanding of quantum science themselves but also to find relatable and age-appropriate ways to introduce these concepts to their students. The dual focus on foundational knowledge and pedagogical strategies suggests a comprehensive approach to incorporating quantum science into elementary education, emphasizing both content knowledge and innovative teaching methodologies.

\subsection{Teachers' Knowledge of Quantum }
Most participants had limited exposure to quantum concepts prior to their work on the project. For many, their knowledge of quantum mechanics was derived from cultural references such as the TV shows ``Big Bang Theory" and ``Quantum Leap," or through Marvel movies rather than formal educational or scientific sources. Two of the participants reported personal connections with quantum scientists, yet they candidly admitted that their grasp of the subject remained on a surface level. One teacher's comment underscored this point: “I know quantum and physics go together in some contexts, but that's about it.” Others expressed a mixture of curiosity and confusion: “I think it's confusing and I think that there's a lot of potential there, but the understanding is hard,” and “That is the first thing [Ant-Man] and then I know it has something to do with the position of electrons... all this happened with quantum technology... I know quantum computing too, and it's a little bit about stuff on small scales.” Another participant reflected on an incidental exposure to the concept through personal conversations, revealing the incidental nature of their learning: “I think also having my husband working in space and physics-related kinds of things, I tend to hear some of those buzzwords but even when he and I were talking about quantum the other day, and I did totally bring it up because of this project, not because this is a normal conversation in our house.” These comments highlighted a common theme: there is a recognition among educators of the importance of quantum concepts and their potential applications, yet there is a clear need for more structured educational resources and professional development opportunities to bridge their knowledge gaps. 

\subsection{Elementary Students' Knowledge of Quantum}
During the project, several participants conducted informal inquiries with their Grade 5 students to discover what their elementary students know about quantum.  We note that during year 1, teachers were not implementing quantum into their classrooms, as our team was still developing resources. De-identified student responses were shared with the researchers and other teachers to inform curriculum development. 

Three teachers asked students to share what they knew about quantum, and their students' responses varied widely. A subset of students demonstrated a basic understanding of quantum science, describing it with terms like ``electro," ``energy," ``physics on a deeper molecular level," ``a small object or atom that scientist can find," and one student used the idea from Schrödinger’s cat thought experiment as ``alive or dead." These responses suggest some students have exposure to concrete quantum concepts. Another group of students thought about quantum in more abstract terms, labeling it as ``advanced science," ``invisible," ``complicated," and encompassing both ``micro and macro" scales. These descriptors indicate a perception of quantum science as a complex and elusive field, highlighting an intuitive grasp of the scales and complexities involved in quantum physics.  Lastly, there were students who associated quantum with a variety of broadly scientific concepts, calling it ``a fancy science word," and linking it to ``space," ``amount of universe," ``solar system," ``black hole," ``gravity," ``galaxy,"  and the ``periodic table,"  Some students are still exploring, they say “quantum and quantity sound the same,” “a type of bug that affects computers, really dangerous to everyone," ``a plant or a animal," , and ``a kind of source to help a kind of sickness." These responses revealed how students often conflate quantum science with other areas of scientific study, suggesting a mix of curiosity and confusion about where quantum physics fits within the broader scientific world. We also noted that many students mentioned they got to know quantum concepts from Marvel movies.

\section{Discussion and Conclusion}
Our findings indicate that elementary students know little about quantum, which is what we expected to find. However, some students already knew a little about quantum, and other students' responses suggested they had developed misconceptions about quantum from popular media. Introducing quantum concepts within elementary STEM classrooms can raise quantum awareness and give students a head-start on developing an understanding of these complex concepts.  Our teachers reported that elementary students were engaged and energized by even the idea of learning about quantum. In this work we aim to spark students' excitement and interest in quantum topics, applications, and careers, which is a fundamental goal of K-12 quantum \cite{b6}.

In our prior work with elementary, middle, and high school teachers \cite{b14}, we found that teachers were excited and interested in quantum, but were concerned about their readiness to learn quantum concepts and their students' ability to learn quantum.  In our previous work we also found that teachers saw the potential to advance equity in STEM through teaching quantum.  Our preliminary findings from year one support these findings.  

All participants recognize their role as STEM educators primarily involves facilitating the development of students' problem-solving skills through scaffolding. They believe that hands-on activities are the most effective method for helping students grasp scientific concepts. There is a consensus among participants that science education involves discovery, risk-taking, and problem-solving. Moreover, it encourages students to engage in collaborative work, fostering a team-oriented approach to learning. This shared understanding of science is crucial as educators look to integrate these principles into classroom activities and curricula. The next stage of this project is to introduce elementary students to quantum concepts in a comprehensible and engaging manner. Many educators are considering the use of illustrated books and interactive games as tools to captivate students' interests and explain abstract concepts to them. As the educators on our team collaboratively develop new resources for elementary school, they are working to make learning more interactive while also demystifying complex quantum concepts so that they are accessible to young learners.

\end{document}